\documentclass[11pt,a4paper,nofootinbib]{article} 
\usepackage{jheppub}
\usepackage{enumerate}
\usepackage{epsfig}
\usepackage{amsmath} 
\usepackage{capt-of}
\usepackage{wasysym} 
\usepackage{mathrsfs} 
\usepackage{graphicx} 
\usepackage{amsfonts} 
\usepackage{amsbsy} 
\usepackage{amscd}
\usepackage{rotating}
\usepackage{slashed, latexsym, verbatim, cancel} 
\usepackage[utf8]{inputenc}
\usepackage {multirow}
\usepackage{color}
\definecolor{myred}{rgb}{0.6,0,0} 
\definecolor{myblue}{rgb}{0,0.2,0.4}
\definecolor{mygreen}{rgb}{0,0.9,0.1}
\definecolor{hc}{rgb}{.9,0.1,0.7}
\definecolor{hcout}{rgb}{.9,0.7,0.9}
\definecolor{Orange}{rgb}{1.,0.65,0.}

\makeatletter

\newcommand{\fmslash}[2][0mu]{%
  \mathchoice
    {\fmsl@sh\displaystyle{#1}{#2}}%
    {\fmsl@sh\textstyle{#1}{#2}}%
    {\fmsl@sh\scriptstyle{#1}{#2}}%
    {\fmsl@sh\scriptscriptstyle{#1}{#2}}}
\newcommand{\fmsl@sh}[3]{%
  \m@th\ooalign{$\hfil#1\mkern#2/\hfil$\crcr$#1#3$}}
\makeatother

\newcommand{\lsim}{{\;\raise0.3ex\hbox{$<$\kern-0.75em\raise-1.1ex\hbox{$\sim$}}\;}}
\newcommand{\gsim}{{\;\raise0.3ex\hbox{$>$\kern-0.75em\raise-1.1ex\hbox{$\sim$}}\;}}

\usepackage{array}
\newcolumntype{C}[1]{>{\centering\arraybackslash$}p{#1}<{$}}

\usepackage{bm} 

\usepackage{cleveref}
\newcommand{\be}{\begin{equation}}
\newcommand{\ee}{\end{equation}}
\newcommand{\bes}{\begin{equation*}}
\newcommand{\ees}{\end{equation*}}
\newcommand{\bea}{\begin{eqnarray}}
\newcommand{\eea}{\end{eqnarray}}
\newcommand{\beas}{\begin{eqnarray*}}
\newcommand{\eeas}{\end{eqnarray*}}

\newcommand{\la}{\lambda}

\newcommand{\wt}{\widetilde}

\title{Electron EDM and Muon anomalous magnetic moment  in Two-Higgs-Doublet Models }
\author{Eung Jin Chun,}
\author{Jongkuk Kim,}
\author{Tanmoy Mondal} 
\affiliation{Korea Institute for Advanced Study, Seoul 02455, Korea}
\emailAdd{ejchun@kias.re.kr}
\emailAdd{tanmoy@kias.re.kr}
\emailAdd{jkkim@kias.re.kr}
  
\abstract{ The CP violating two-Higgs doublet model of type-X may enhance significantly the electric and magnetic 
moment of leptons through two-loop Barr-Zee diagrams. We analyze the general parameter space of the type-X 2HDM 
consistent with the muon $g-2$ and the electron EDM measurements to show how strongly the CP violating parameter 
is constrained in the region explaining the muon $ g-2$ anomaly.
}

\preprint{KIAS-P19032}
\date{\today}


\begin{document}
\maketitle


\section{Introduction}
After the discovery of the 125 GeV Higgs boson \cite{lhc2012,Chatrchyan:2012xdj}, its nature has been found to follow very closely to the prediction 
of the Standard Model (SM) in terms of its production and decay properties \cite{atlas-higgs} as well as the CP property 
\cite{cms-higgs}. Nonetheless, we expect that there must be new physics beyond SM for various theoretical and phenomenological reasons.
Extending the Higgs sector with an additional doublet is an interesting option which has potentially important implications on 
several new physics phenomena. A Two-Higgs Doublet Model (2HDM)  is able to realize electroweak baryogenesis  
\cite{turok90}, and the relevant CP violation  (CPV) may appear in electric dipole moments (EDMs) of fermions \cite{Barroso:2012wz,jung13,ipek13,Cheung:2014oaa,inoue14,Shu:2013uua,Keus:2017ioh}.
Such CP violation effect could be probed as well in the future collider experiments \cite{Chen:2015gaa,chen17,fontes17,aoki18}.

On the other hand, the observed deviation of the muon anomalous moment \cite{bnl}
can be explained in type-X 2HDM \cite{broggio14,jinsu16,Cao:2009as,Wang:2014sda,Ilisie:2015tra,Abe:2015oca,Cherchiglia:2017uwv,Wang:2018hnw}. Both for EDM and the muon $g-2$, 
sizable contributions come from two-loop Barr-Zee (BZ) diagrams \cite{barr-zee}. However, the favored parameter spaces are largely orthogonal to each other.  
The $(g-2)_\mu$  data can be accommodated only by type-X 2HDM with large $\tan\beta$ and a very light pseudoscalar, while strong 
phase transition for electroweak baryogenesis can be realized for a very heavy pseudoscalar Higgs boson preferring low $\tan\beta$ \cite{dorsch16}. 
A recent study \cite{Wang:2018hnw} showed that strongly first order phase transition could be obtained for some limited parameter region explaining the muon $g-2$ deviation with a light pseudoscalar.
It is interesting to see whether sizable CPV can be allowed for successful electroweak baryogenesis.
Collider searches for such a light pseudoscalar at LHC were studied in \cite{Chun:2015hsa,Chun:2017yob,Chun:2018vsn}.
However, more studies are needed to probe the whole parameter space favorable for the muon $g-2$.

In this paper, we analyze the general (complex) parameter region of the type-X 2HDM compatible with the EDM and
$(g-2)_\mu$ measurements. Since both quantities come from the same type of loop diagrams, 
only the electron EDM is highly enhanced in the region compatible with the muon $g-2$ anomaly and thus 
the CP violating coupling is severely constrained by the electron EDM limit.   

The contents of the paper are as follows. 
In Section \ref{Sec:CP}, we give a brief introduction to the general properties of the CP-violating type-X 2HDM, and 
define our input parameter set.
In Section \ref{ConstOn2HDM}, we describe all the theoretical and experimental bounds on the parameter space to find allowed region 
and the results of analysis are presented in Section \ref{results}. Finally we  conclude in Section \ref{conclusions}.   
In Appendix, we collect all the necessary formula to compute EDM in 2HDM.

\section{The $CP$ violating type-X two-Higgs Doublet Model}\label{Sec:CP}

In this section we will briefly summarize the CP violating two-Higgs doublet model. The model consists of two 
scalar doublets $\Phi_1$ and $\Phi_2$ of equal hypercharge. The general form of the scalar potential is given by
\begin{eqnarray}
\nonumber V_{\mathrm{2HDM}} &=& -m_{11}^2\Phi_1^{\dagger}\Phi_1 - m_{22}^2\Phi_2^{\dagger}\Phi_2 -\Big[m_{12}^2\Phi_1^{\dagger}\Phi_2 + \mathrm{h.c.}\Big]
+\frac{1}{2}\lambda_1\left(\Phi_1^\dagger\Phi_1\right)^2+\frac{1}{2}\lambda_2\left(\Phi_2^\dagger\Phi_2\right)^2 \\
\nonumber && +\lambda_3\left(\Phi_1^\dagger\Phi_1\right)\left(\Phi_2^\dagger\Phi_2\right)+\lambda_4\left(\Phi_1^\dagger\Phi_2\right)\left(\Phi_2^\dagger\Phi_1\right)
+\Big\{ \frac{1}{2}\lambda_5\left(\Phi_1^\dagger\Phi_2\right)^2+\Big[\lambda_6\left(\Phi_1^\dagger\Phi_1\right) \\
&& +\lambda_7\left(\Phi_2^\dagger\Phi_2\right)\Big]\left(\Phi_1^\dagger\Phi_2\right) + \rm{h.c.}\Big\}.
\label{eq:2hdm-pot}
\end{eqnarray}
In general fermions can couple to both the scalars  and flavor changing neutral current interactions (FCNCs) can appear 
at tree level. To avoid the problematic FCNCs we impose a $\mathbb{Z}_2$ symmetry under which the scalars are oppositely 
charged, $\Phi_1\rightarrow-\Phi_1$ and $\Phi_2\rightarrow \Phi_2$, and the fermions are also charged appropriately.
 This $\mathbb{Z}_2$ charge assignment forbids the 
$\lambda_6$ and $\lambda_7$ term in the scalar potential in Eq.~\ref{eq:2hdm-pot}. Here, we include  the soft $\mathbb{Z}_2$ breaking term $m_{12}^2$ which can generates CP violation in the scalar sector~\cite{Accomando:2006ga}. 
This term is also important to keep the quartic coupling $\lambda_1$ below perturbativity limit~\cite{Gunion:1989we,Gunion:2002zf}. 
The complex parameters in the potential are $m_{12}^2$ and $\lambda_5$ while all other parameters are real. The scalars can be parameterized as 
\begin{eqnarray}\label{eq:scalars-param}
\Phi_1=\begin{pmatrix}
\phi_1^+ \\
\frac{1}{\sqrt2} (v_1 + \phi_1^{0\,r} + i \phi_1^{0\,i})
\end{pmatrix}, \ \ 
\Phi_2=\begin{pmatrix}
\phi_2^+ \\
\frac{1}{\sqrt2} (v_2 + \phi_2^{0\,r} + i \phi_2^{0\,i})
\end{pmatrix} \ ,
\end{eqnarray}
where the vacuum expectation values ($vev$s) $v_1, v_2$ are real and $v_1^2 + v_2^2 = v^2 = \left(246 \rm{ GeV}\right)^2$. 
In general one of the $vev$s can be complex. Under the global phase transformation the couplings $m_{12}^2$ and $\lambda_5$ can absorb the global phase. Thus $v_1,$ and  $v_2$ are real parameters.
Minimization of the scalar potential yields the following relations (we define $\tan\beta = v_2 \// v_1$), 
\begin{eqnarray}\label{eq:minimization}
m_{11}^2 &=& \lambda_1 v^2 \cos^2\beta + (\lambda_3 + \lambda_4 + {\rm Re} (\lambda_5)) v^2 \sin^2\beta - {\rm Re} (m_{12}^2) \tan\beta \ ,
\label{eq:mini1}\\
m_{22}^2 &=& \lambda_2 v^2 \sin^2\beta + (\lambda_3 + \lambda_4 + {\rm Re} (\lambda_5)) v^2 \cos^2\beta - {\rm Re} (m_{12}^2) \cot\beta \ ,
\label{eq:mini2}\\
{\rm Im} (m_{12}^2 )&=&v^2 \sin\beta\cos\beta \;{\rm Im} ( \lambda_5 ) \ . \label{eq:mini3} 
\end{eqnarray}
From Eq.~\ref{eq:mini3} it is evident that there is only one free CPV parameter and from now on we will use $\rm{Im}(\lambda_5)$ as free 
parameter to quantify the CP violation. 

\subsection{Scalar Spectrum}
After the electroweak symmetry breaking (EWSB) we obtain one physical charged scalar  and one Goldstone boson,
\be
\begin{pmatrix}
G^+ \\
h^+
\end{pmatrix} = \begin{pmatrix}
\cos\beta & \sin\beta \\
-\sin\beta&\cos\beta
\end{pmatrix} 
\begin{pmatrix}
\phi_1^+ \\
\phi_2^+
\end{pmatrix},
\ee
and mass of the charged scalar reads as, 
\be
m_{h^\pm}^2 =\frac{{\rm Re} (m_{12}^2)}{2\,\cos\beta \sin\beta} -\frac12 \left(\la_4+ {\rm Re} (\lambda_5)\right)v^2 = \frac12 \left(2\nu - \lambda_4 - {\rm Re}(\lambda_5) \right)  v^2,
\ee
where we define,
$$
\nu \equiv \frac{{\rm Re} (m_{12}^2)}{{2\,\,v^2 \cos\beta \sin\beta}}.
$$
Similarly, mixing between the two CP-odd components of the scalar doublets generates a Goldstone and a CP-odd scalar,
\be
\begin{pmatrix}
G^0 \\
A^0
\end{pmatrix} = \begin{pmatrix}
\cos\beta & \sin\beta \\
-\sin\beta&\cos\beta
\end{pmatrix} 
\begin{pmatrix}
\phi_1^{0\,i} \\
\phi_2^{0\,i}
\end{pmatrix}.
\ee
In the case of CP conserving 2HDM model the combination $A^0$ is the mass eigenstate. Since we have explicit 
CP violation, $A^0$ will further mix with $\phi_1^{0\,r}$ and $\phi_2^{0\,r}$ to produce three neutral scalar mass eigenstates. 
The mass matrix 
is given by,
\be
\mathcal{L}_{mass} = \frac12
\begin{pmatrix} \phi_1^{0\,r} & \phi_2^{0\,r} & A^0\end{pmatrix} \mathcal{M}^2 \begin{pmatrix} \phi_1^{0\,r} \\ \phi_2^{0\,r} \\ A^0\end{pmatrix}.
\ee
The neutral mass squared matrix is,
\be\label{eq:mass_sq}
\mathcal{M}^2 = v^2 \begin{pmatrix}
\lambda_1c_\beta^2+\nu s_\beta^2 & (\lambda_{345} - \nu)c_\beta s_\beta & -\frac{1}{2}{\rm Im}(\lambda_5) \, s_\beta \\
(\lambda_{345} - \nu)c_\beta s_\beta & \lambda_2 s_\beta^2+\nu c_\beta^2 & -\frac{1}{2}{\rm Im}(\lambda_5)\, c_\beta\\
-\frac{1}{2}{\rm Im}(\lambda_5)\, s_\beta & -\frac{1}{2}{\rm Im}(\lambda_5)\, c_\beta & \nu - {\rm Re}(\lambda_5)
\end{pmatrix} \ ,
\ee
where $c_\beta=\cos\beta, s_\beta=\sin\beta$ and $\lambda_{345} = (\lambda_3 + \lambda_4 + {\rm Re} (\lambda_5))$. The mass matrix can be 
diagonalized by a rotation matrix $R$ such that $\mathcal{M}_{diag} = R \mathcal{M}^2 R^T$ and we get,
\be
\mathcal{L}_{mass} = \frac12 \begin{pmatrix} h_1 & h_2 & h_3\end{pmatrix} \mathcal{M}_{diag}^2 \begin{pmatrix} h_1 \\ h_2 \\ h_3\end{pmatrix}, 
\textrm{  where   }   \begin{pmatrix} h_1 \\ h_2 \\ h_3\end{pmatrix} =  R \begin{pmatrix} \phi_1^{0\,r} \\ \phi_2^{0\,r} \\ A^0\end{pmatrix}.
\ee
We define $\mathcal{M}_{diag} = \textrm{diag}(m_{h_1}^2,m_{h_2}^2,m_{h_3}^2)$ and  $h_1$  is  the Higgs boson observed at the LHC with mass $m_{h_1} =$ 125 GeV. As we will see that allowed CPV is small, the mass eigenstate $h_3$ 
is predominantly CP-odd whereas $h_1$ and $h_2$ are mostly CP-even.
The rotation matrix $R$ depends on three mixing angles,
$\alpha,~\alpha_b$ and $\alpha_c$ where the last two angles stem from CP violation. The matrix $R$ can be parameterized as 
\begin{eqnarray}
R 
&=&\begin{pmatrix}
-s_{\alpha}c_{\alpha_b} & c_{\alpha}c_{\alpha_b} & s_{\alpha_b} \\
s_{\alpha}s_{\alpha_b}s_{\alpha_c} - c_{\alpha}c_{\alpha_c} & -s_{\alpha}c_{\alpha_c} - c_{\alpha}s_{\alpha_b}s_{\alpha_c} & c_{\alpha_b}s_{\alpha_c} \\
s_{\alpha}s_{\alpha_b}c_{\alpha_c} + c_{\alpha}s_{\alpha_c} & s_{\alpha}s_{\alpha_c} - c_{\alpha}s_{\alpha_b}c_{\alpha_c} & c_{\alpha_b}c_{\alpha_c}
\end{pmatrix} ,
\end{eqnarray}
where $s_{\alpha} = \sin\alpha$ etc and $-\dfrac{\pi}{2}\leq \alpha,\alpha_b,\alpha_c \leq \dfrac{\pi}{2}$.

There are ten real parameters in the scalar potential which are,
\be
\la_1,\,\la_2,\,\la_3,\,\la_4,\,{\rm Re} (\lambda_5),\, {\rm Im} (\lambda_5),\,m_{11}^2,\,m_{22}^2,\, {\rm Re} (m_{12}^2), \;\;\; \&\;\; {\rm Im} (m_{12}^2).
\ee
It is possible to re-express the parameters of the potential in terms of the following phenomenological parameters,
\be\label{eq:pheno_param_old}
m_{h_1},\,m_{h_2},\,m_{h_3},\,m_{h^\pm},\,v ({\rm vev}),\, \tan\beta,\, \alpha,\,\alpha_b,\, \alpha_c,\,  \;\;\; \&\;\; \nu.  
\ee
We can express the quartic couplings in terms of the the phenomenological parameters~\cite{inoue14} : 
\begin{subequations}\label{eq:lambdas}
\begin{eqnarray}
\lambda_1 &=& \frac{m_{h_1}^2 R_{11}^2 + m_{h_2}^2 R_{21}^2 + m_{h_3}^2 R_{31}^2}{v^2 \cos\beta^2} - \nu \tan^2\beta \ , \\
\lambda_2 &=& \frac{m_{h_1}^2 R_{21}^2 + m_{h_2}^2 R_{22}^2 + m_{h_3}^2 R_{32}^2}{v^2 \sin\beta^2} - \frac{\nu}{\tan^2\beta} \ , \\
{\rm Re}(\lambda_5) &=& - \frac{m_{h_1}^2  R_{31}^2 + m_{h_2}^2 R_{32}^2 + m_{h_3}^2 R_{33}^2}{v^2} + \nu\ , \\
\lambda_4 &=& 2 \nu - {\rm Re}(\lambda_5) - \frac{2 m_{H^+}^2}{v^2} \ , \\
\lambda_3 &=& \nu - \frac{m_{h_1}^2 R_{11}R_{12} - m_{h_2}^2 R_{21}R_{22} - m_{h_3}^2R_{31}R_{32}}{v^2\sin\beta\cos\beta} - \lambda_4 - {\rm Re}(\lambda_5 ).
\end{eqnarray}
\end{subequations}

Note that all the parameters in Eq.~\ref{eq:pheno_param_old} are not independent since there is only one independent CP violating parameter as indicated by Eq.~\ref{eq:mini3}. For simplicity we have used ${\rm Im} (\lambda_5)$ as the measure 
of CP violation in our model and computed $\alpha_b$ and $\alpha_c$ accordingly. Moreover, the angle $\alpha$ is re-expressed by using  $\cos(\beta-\alpha)$ where the angle $(\beta-\alpha)$ diagonalizes the CP even neutral scalar mass  matrix in the CP conserving 2HDM scenario. Consequently, the phenomenological parameter set we have used for parameter scan is : 
\be\label{eq:pheno_param}
m_{h_1},\,m_{h_2},\,m_{h_3},\,m_{h^\pm},\,v ({\rm vev}),\, \tan\beta,\, \cos(\beta-\alpha),\,\,  {\rm Im} (\lambda_5) \;\;\; \&\;\; \nu.
\ee

To compute the quartic couplings in Eq.~\ref{eq:lambdas} we need to recast the angles $\alpha_b$ and $\alpha_c$ in terms of the parameters as 
shown in Eq.~\ref{eq:pheno_param}. From the diagonalization of the neutral scalar mass matrix we have, $\mathcal{M}^2 = R^T \mathcal{M}_{diag}^2 R $ and using Eq.~\ref{eq:mass_sq} we get the following relations,
\be
\left(R^T \mathcal{M}_{diag}^2R\right)_{13} = -\frac{1}{2}{\rm Im}(\lambda_5) \, s_\beta , \  \ \& \  \ 
\left(R^T \mathcal{M}_{diag}^2R\right)_{23} =-\frac{1}{2}{\rm Im}(\lambda_5)\, c_\beta.
\ee
Assuming small CP violation we can solve the above two equations under the assumption that $\sin^2\alpha_b,\ \sin^2\alpha_c \to 0$. We obtain the following expression up to leading order of $ {\rm Im} (\lambda_5)$,
\bea\label{eq:mixing_angles}
\sin\alpha_b \simeq- \frac{v^2 \cos(\alpha+\beta)} {2\ (m_{h_1}^2-m_{h_3}^2)} \ \ {\rm Im} (\lambda_5) \nonumber \\
\sin\alpha_c \simeq \frac{v^2 \sin(\alpha+\beta)} {2\ (m_{h_2}^2-m_{h_3}^2)} \ \ {\rm Im} (\lambda_5).
\eea
These expressions are valid as long as the mass of $h_3$ is not very close to any of the other two scalars. For our numerical computation we 
have used the full expressions without any assumptions. Now using our phenomenological parameter set we can compute all the scalar 
mixing and quartic couplings.

\subsection{Yukawa and Gauge Interactions of the neutral scalars}
The interaction of the fermions with the two scalar doublets depends on the $\mathbb{Z}_2$ symmetry we invoked to remove the tree 
level FCNCs. Depending on the $\mathbb{Z}_2$ charge assignment for fermions, four independent types of Yukawa interactions are 
allowed~\cite{Branco:2011iw}.  We are interested in the scenario where the leptons couple to one doublet ($\Phi_1$) and the quarks 
couple to the other doublet ($\Phi_2$). This model is known as type-X 2HDM or leptophilic 2HDM. The Yukawa terms read as,
\be\label{eq:yukawa}
-{\cal L}_Y= Y^u\bar{ Q_L} \wt \Phi_2 u_R + Y^d  \bar{ Q_L} \Phi_2 d_R+Y^e\bar{ L_L} \Phi_1 \ell_R + h.c..
\ee
The above Lagrangian can be realized if $\ell_R$ is odd under the $\mathbb{Z}_2$ symmetry  while the other fermions are even under it.
After the EWSB the general form of the Yukawa interaction is,
\be
\mathcal{L}_Y = -\frac{m_f}{v} h_i \left( c_{f,i} \bar f f + i\ \tilde c_{f,i} \bar f \gamma_5 f  \right),
\ee
where $ c_{f,i}$ and $\tilde c_{f,i}$ are the Yukawa modifiers. We can express the Yukawa modifiers in terms of mixing matrix($R$) and 
$\tan\beta$ as shown in Table~\ref{tab:yukawa_mod} where $ c_{u,i}$ is the universal modifier for all generations of up type quarks 
and same is true for down type quarks ($c_{d,i}$), and leptons ($ c_{e,i}$). In the limit $\cos(\beta-\alpha) \to 0 $, the modifiers to the SM like Higgs 
$ c_{f,1}$ goes to +1 essentially restoring the SM Yukawa coupling. This is called right sign (RS) Yukawa limit. However, if $\cos(\beta-\alpha)$ takes the particular 
value $2/\tan\beta$ then the lepton Yukawa modifier $c_{e,1}$  becomes `-1' and is known as wrong sign (WS) Yukawa limit.
\begin{table}[t]
\begin{center}
\begin{tabular}{|c||c|c|c|c|c|c||c|}
\hline
& $ c_{u,i}$ & $ c_{d,i} $ & $ c_{e,i}$ & $\tilde c_{u,i}$ & $\tilde c_{d,i}$ & $\tilde c_{e,i}$& $a_i$ \\ \hline
Type-X & $\dfrac{R_{i2}}{\sin\beta}$ & $\dfrac{R_{i2}}{\sin\beta}$ & $\dfrac{R_{i1}}{\cos\beta}$ & - $\dfrac{R_{i3}}{\tan\beta}$  & $\dfrac{R_{i3}}{\tan\beta}$ & - $ R_{i3} \tan\beta$ & $R_{i1}\cos\beta + R_{i2}\sin\beta $ \\
 \hline
\end{tabular}
\end{center}
 \caption{The multiplicative factors of Yukawa interactions in type-X 2HDM}
 \label{tab:yukawa_mod}
\end{table}

The couplings between neutral Higgs bosons and the gauge bosons can be written as,
\be
\mathcal{L}_{gauge} = \dfrac{h_i}{v}a_i\ \left(2m_W^2\ W_\mu W^\mu + m_Z^2\ Z_\mu Z^\mu  \right),
\ee
where the expression for $a_i$ is shown in Table~\ref{tab:yukawa_mod}. 


\section{Theoretical and experimental constraints } \label{ConstOn2HDM}

\subsection{Perturbativity and vacuum stability}

The quartic couplings have to satisfy the theoretical conditions for vacuum stability and perturbativity. 
For perturbativity  we ensure that $|\la_i| < 4 \pi$. The vacuum stability conditions are~\cite{Gunion:2002zf},
\begin{eqnarray}
0<\lambda_1,~\la_2 < 4\pi,  \ \lambda_3 > - \sqrt{\lambda_1 \lambda_2}, \ \ \& \ \ \lambda_3 + \lambda_4 - |\lambda_5| > - \sqrt{\lambda_1 \lambda_2} \ .
\label{eq:vacpert}
\end{eqnarray}
The limit on the quartic coupling and consequently on the scalar masses depends on the parameter $\cos(\beta-\alpha)$.
To satisfy  the perturbativity and vacuum stability constraints the heavy scalar mass need to obey the following limits~\cite{broggio14,Wang:2014sda}, 
\bea
m_{h_2} \simeq m_{h^\pm} &\leq& 250 \textrm{ GeV} \hspace{1cm} \textrm{(RS scenario)}\nonumber \\
m_{h_2} \simeq m_{h^\pm} &\leq& \sqrt{\la_{max}}\ v = \sqrt{4\pi}\ v\hspace{1cm} \textrm{(WS scenario)}.
\eea 
We will appropriately choose the value of $\cos(\beta-\alpha)$  and $\nu$ depending on the charged Higgs mass to satisfy these conditions.

\subsection{Muon anomalous magnetic moment}

The muon anomalous magnetic moment is a long standing puzzle with disagreement between the SM prediction 
 and the experimental determination. 
The recent estimate of theoretical~\cite{Keshavarzi:2018mgv} and experimental~\cite{bnl}  value for $a_\mu$ is:
\bea
a^{\rm th}_\mu &=& \left( 116591820.4 \pm 35.6 \right) \ \times 10^{-11}, \\
a^{\rm ex}_\mu  &=& \left( 116592089 \pm 63 \right)  \times 10^{-11}. \nonumber
\eea
This gives us  a $3.7\sigma$ deviation: 
\begin{eqnarray}  \label{eq:delta-amu}
\delta a_\mu &\equiv& a^{\rm ex}_\mu - a^{\rm th}_\mu = \left( 270.6 \pm 72.6\right)\times 10^{-11}.
\end{eqnarray}
The Fermilab Muon $g-2$ experiment (E989) is aiming to measure $a_\mu$ with a relative uncertainty of 140 ppb which has the potential to confirm the discrepancy 
with a $7\sigma$ significance~\cite{fermilab} assuming that the central values of $a^{\rm th}_\mu$ and $a^{\rm ex}_\mu$ remain the same. 
In the CP-conserving  type-X 2HDM,  such a deviation can be explained by enhanced contribution coming from the two loop Barr-Zee diagram for large $\tan\beta$ and low pseudoscalar mass.  In the presence of CP violation, there will be extra  contributions 
from the  CPV Yukawa couplings. However the additional contribution is negligible as we will see that 
the EDM bound strongly  suppresses the CPV coupling  (${\rm Im}(\lambda_5)$).

\subsection{Electron electric dipole moment}

\begin{figure}[t]
\begin{center}
  \includegraphics[width=12cm,angle=0]{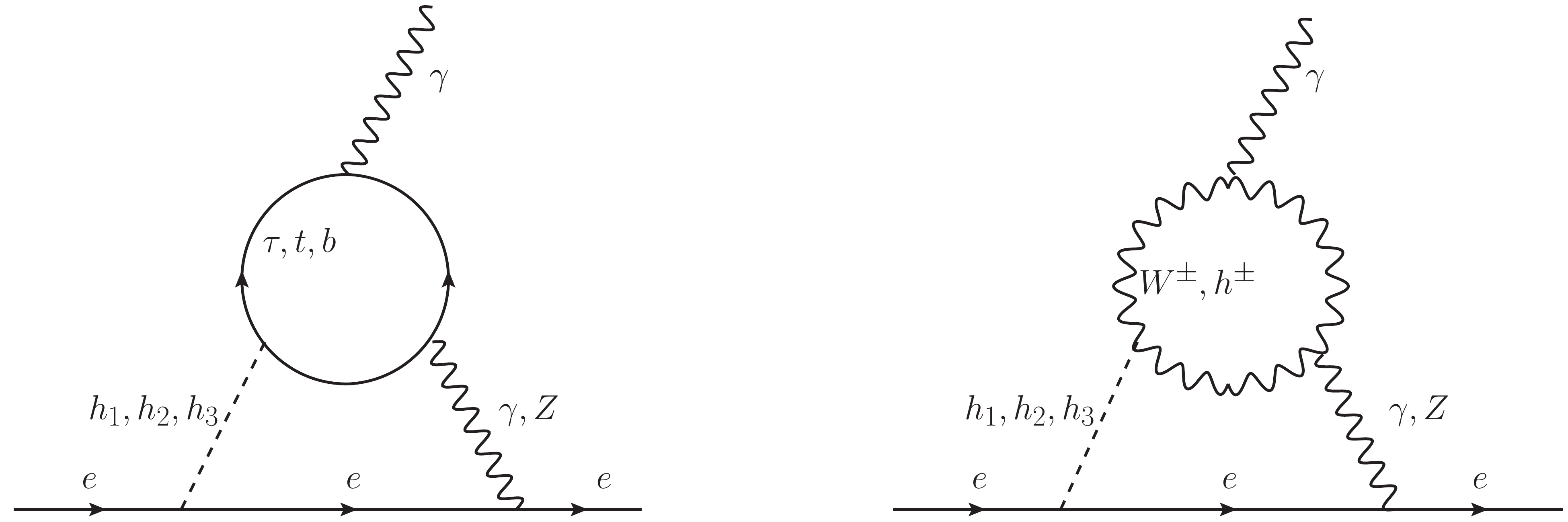}
  \includegraphics[width=12cm,angle=0]{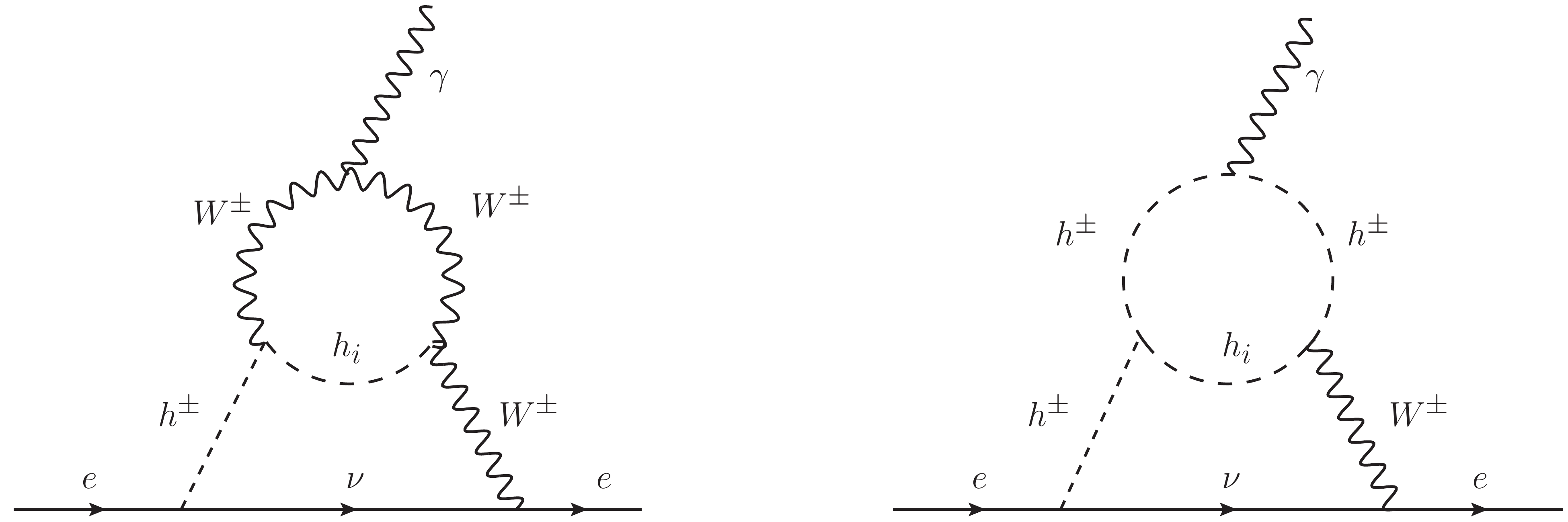}
  \caption{Representative 2-loop Barr-Zee diagrams which generate EDM due to CP violation. 
  Similar diagrams contribute dominantly in muon anomalous magnetic moment when electron 
  is replaced by muon.}
 \label{fig:BZ_diagram}
\end{center}
\end{figure}

EDMs are very sensitive probes of new physics that contains CP-violating phases. 
In our model the complex quartic coupling in the Higgs sector  results in CP violating phases in Yukawa couplings  
as shown in Eq.~\ref{eq:yukawa}.
Note that electron EDM can come from both one loop and two loop diagrams. Contribution from 1-loop diagrams is induced by the neutral Higgs bosons or the charged Higgs. 
This contribution is proportional to the third power of the electron Yukawa 
coupling and is negligible. The other contribution originates from the so called Barr-Zee diagrams as shown in Fig.~\ref{fig:BZ_diagram}. This might provide sizable EDM since they are proportional to one power of the 
electron Yukawa coupling. 

The effective dipole-moment operator can 
be written as
\be
\mathcal{L}_{eff} = - \frac{i}{2} d_e\ (\bar {\psi_e}\sigma_{\mu\nu}\gamma_5\ \psi_e) F^{\mu\nu}
\equiv - i \frac{\delta_e}{v^2} m_e e \ (\bar {\psi_e}\sigma_{\mu\nu}\gamma_5\ \psi_e) F^{\mu\nu}
\ee
where we have used the Higgs vacuum expectation value $v= 246$ GeV as the cut-off scale for new physics. 
We have summarized all the 2-loop contributions to the coefficient $\delta_e$  to compute the EDM in Appendix~\ref{app:wilson}. 

The current upper bound on electron EDM is reported by the ACME Collaboration ~\cite{Andreev:2018ayy}:
\begin{eqnarray}
\vert d_e \vert &<& 1.1\times 10^{-29} {\ \rm e~cm},
\end{eqnarray}
at 90\% confidence level.
We will use this value to constrain the amount of CPV allowed in this model.

\subsection{Lepton universality constraints}

The test of the lepton universality has been evaluated from $Z$ decay and $\mu/\tau$ decays at the level of $0.1\%$~\cite{ALEPH:2005ab,Amhis:2016xyh}.
In the type-X 2HDM, large loop corrections to the lepton universality can arise due to the enhanced leptonic couplings of extra Higgs bosons at large $\tan\beta$. 
Such effects were well studied in Refs.~\cite{Denner:1991ie,Krawczyk:2004na,jinsu16}.

Let us first consider the test of the lepton universality by SLD and LEP experimental data \cite{ALEPH:2005ab}.
The obtained Z-decay data can be converted into the leptonic branching ratios which are given by
\begin{eqnarray}
\frac{ \Gamma\left(Z \to \mu^+\mu^- \right) }{ \Gamma\left(Z \to e^+e^- \right) } &=& 1.0009 \pm 0.0028,\nonumber\\
\frac{ \Gamma\left(Z \to \tau^+\tau^- \right) }{ \Gamma\left(Z \to e^+e^- \right) } &=& 1.0019 \pm 0.0032,
\end{eqnarray}
with a positive correlation of $0.63$.
For each lepton flavor, we can calculate different quantities which can be parameterized as
\begin{eqnarray}
\delta_{\ell\ell} &\equiv& \frac{ \Gamma\left(Z \to \ell^+\ell^- \right) }{ \Gamma\left(Z \to e^+e^- \right) }-1.
\end{eqnarray}
In the limit of large $\tan\beta$ with $m_\mu \to 0$, the correction on $\delta_{\tau\tau}$ only exists.

The other lepton universality test taken by HFAG has been measured through the pure leptonic and semi-hadronic decay processes \cite{Amhis:2016xyh}:
\begin{eqnarray} \label{hfag-data}
&&
\left( g_\tau \over g_\mu \right) = 1.0010 \pm 0.0015, \quad
\left( g_\tau \over g_e \right) = 1.0029 \pm 0.0015, \quad
\left( g_\mu \over g_e \right) = 1.0019 \pm 0.0014,  \nonumber\\
&&
\left( g_\tau \over g_\mu \right)_\pi = 0.9961 \pm 0.0027, \quad
\left( g_\tau \over g_\mu \right)_K= 0.9860 \pm 0.0070,
\end{eqnarray}
with the correlation matrix
\begin{equation} \label{hfag-corr}
\left(
\begin{array}{ccccc}
1 & +0.53 & -0.49 & +0.24 & +0.11 \\
+0.53  & 1     &  + 0.48 & +0.26    & +0.10 \\
-0.49  & +0.48  & 1       &   +0.02 & -0.01 \\
+0.24  & +0.26  & +0.02  &     1    &     +0.06 \\
+0.11  & +0.10  & -0.01  &  +0.06  &   1 
\end{array} \right) .
\end{equation}
In the large $\tan\beta$ limit of the type-X 2HDM, two crucial corrections to $\tau$ decays can arise.
One correction comes from the tree-level contribution of the heavy charged Higgs boson and the other stems from the one-loop corrections of the extra Higgs bosons.

Following the $\chi^2-$analysis in Ref.~\cite{jinsu16}, we calculate the theoretical bounds on the lepton universality.

\section{Results} \label{results}

\begin{figure}[t]
\begin{center}
  \includegraphics[width=7cm,angle=0]{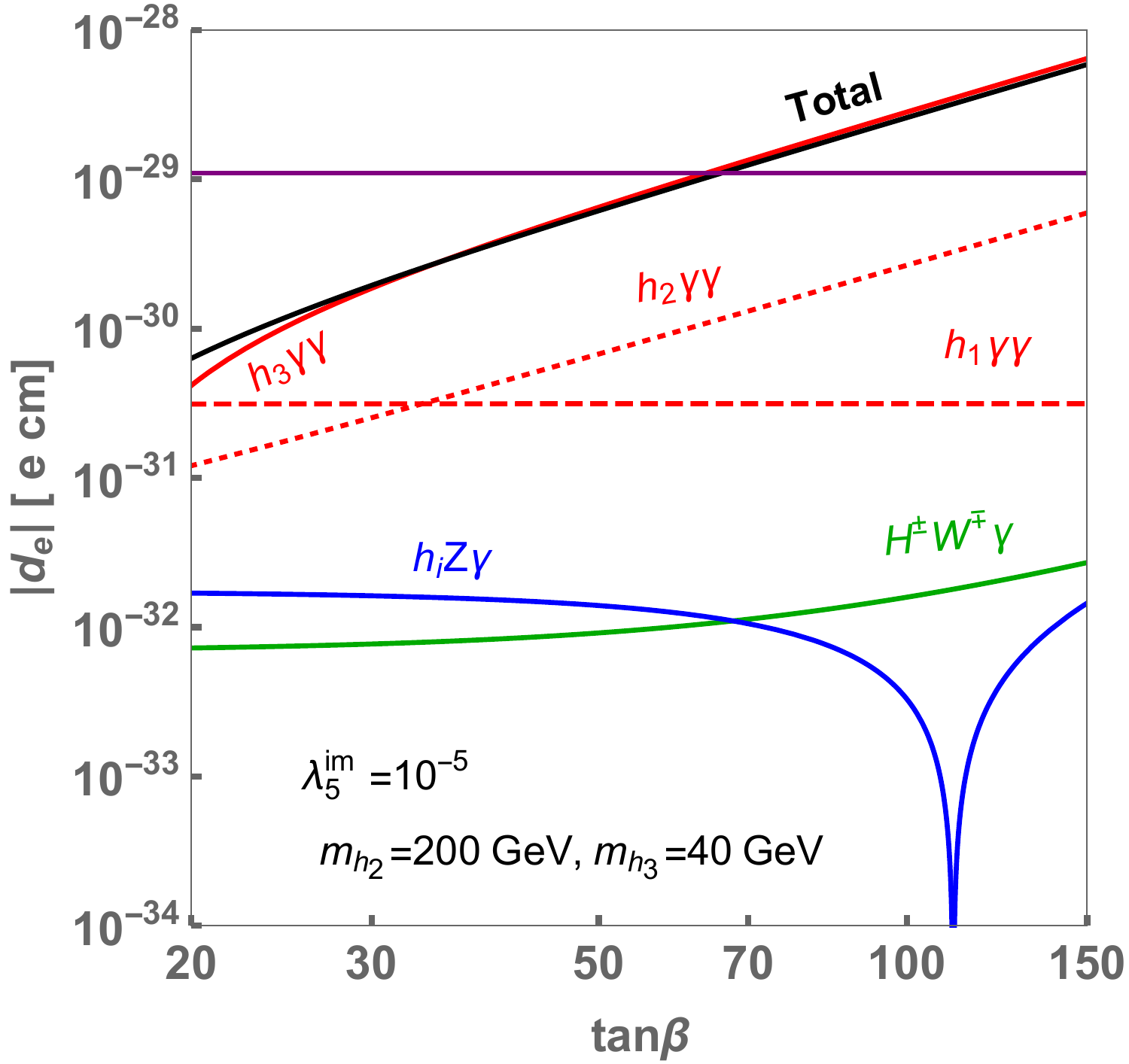}
   \includegraphics[width=7cm,angle=0]{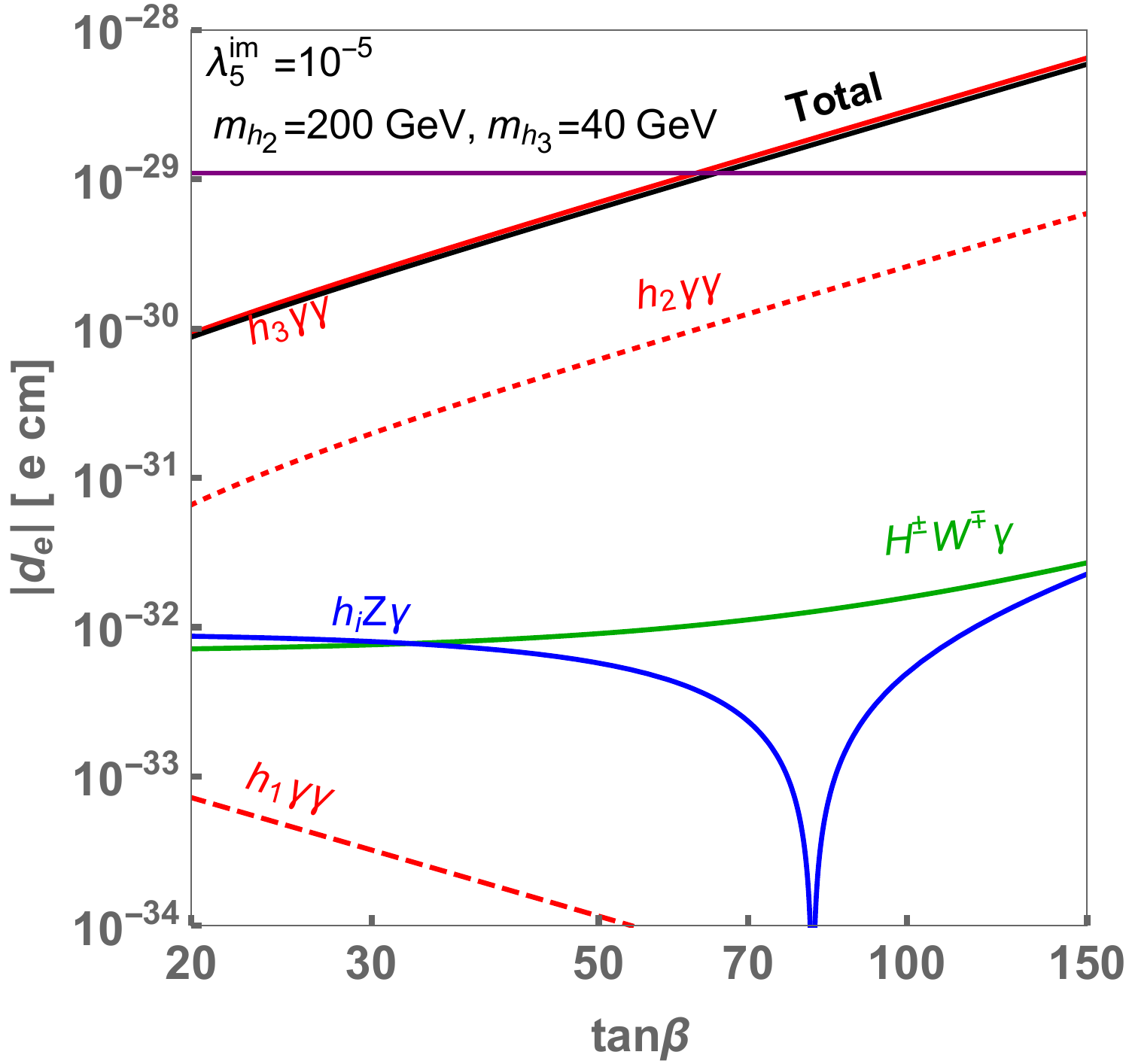}
  \caption{Absolute contribution on electron EDM coming from different Barr-Zee diagrams as a function of $\tan\beta$. We have fixed 
$m_{h_3}=40$ GeV, $m_{h_2}=m_{h^\pm} = 200$ GeV and  ${\rm Im}(\lambda_5) = 10^{-5}$.  In left panel we plot the  RS scenario which corresponds to $\cos(\beta-\alpha)\simeq 0$ and the right panel depicts the WS scenario with $\cos(\beta-\alpha)\simeq 2/\tan\beta$. For both the cases the dominant contribution comes from the $h_3\gamma\gamma$ BZ diagram as shown in Fig.\ref{fig:BZ_diagram}. The parameter ${\rm Im}(\lambda_5)$ is denoted as  $\la_5^{\rm{im}}$ in the plots. }
 \label{fig:edm_separate}
\end{center}
\end{figure}

We are now ready to discuss how the parameter space of type-X 2HDM is limited by  the constraints described in the previous section.
 To satisfy the electroweak precision observable like isospin violation  we assume degeneracy between the heavy scalar $h_2$ and the charged Higgs $h^\pm$  mass keeping  $m_{h_3}$ free. 

In Fig.~\ref{fig:edm_separate} we depict absolute contribution to electron EDM coming from different BZ diagrams in this model for both RS (left panel) and WS scenario(right panel). We have fixed $m_{h_3}=40$ GeV, $m_{h_2}=m_{h^\pm} = 200$ GeV and  ${\rm Im}(\lambda_5) = 10^{-5}$ . 
The total contribution is shown in black curve where dominant contribution is coming from a light $h_3$ mediated diagram with an internal photon line and a tau loop as shown in top left panel of Fig.~\ref{fig:BZ_diagram}. This is due to the fact that  $h_3$ is light and its coupling to leptons is $\tan\beta$ enhanced. Contribution coming from the heavy scalar, $h_2$ is suppressed due to its mass, whereas contributions originate from the BZ diagrams with an internal $Z$ boson are order of magnitude small since $Z$ boson is heavy and the $Z f \bar f$ vector coupling is small. The $h^\pm W^\pm\gamma$ diagrams, 
as shown in lower panel of Fig.~\ref{fig:BZ_diagram} yield sub dominant contributions since both $W$ and $h^\pm$ are relatively heavy. 
The SM Higgs contribution comes from $\sin\alpha_b$ which is proportional to $\cos(\beta+\alpha)$. Hence the SM Higgs contribution goes as $1/\tan\beta\ (1/\tan^3\beta)$ in RS(WS) scenario for large $\tan\beta$. However, for RS scenario there is no $\tan\beta$ dependence since the dominant contribution is generated by the coupling 
$\tilde{c}_{e,1}$ which is proportional to $\tan\beta$(see Table~\ref{tab:yukawa_mod}) and cancels the $\tan\beta$ dependency. The 
$\tan\beta$ dependency in WS limit is obvious. The contribution remains sub dominant due to absence of any $\tan\beta$ enhancement in the SM Higgs mediated BZ diagrams. Since we have plotted the absolute value of EDM, the spike appears when contribution from a particular diagram changes sign. The purple
horizontal bar depicts the present limit on EDM as reported by the ACME Collaboration.

\begin{figure}[t]
\begin{center}
  \includegraphics[width=8cm,angle=0]{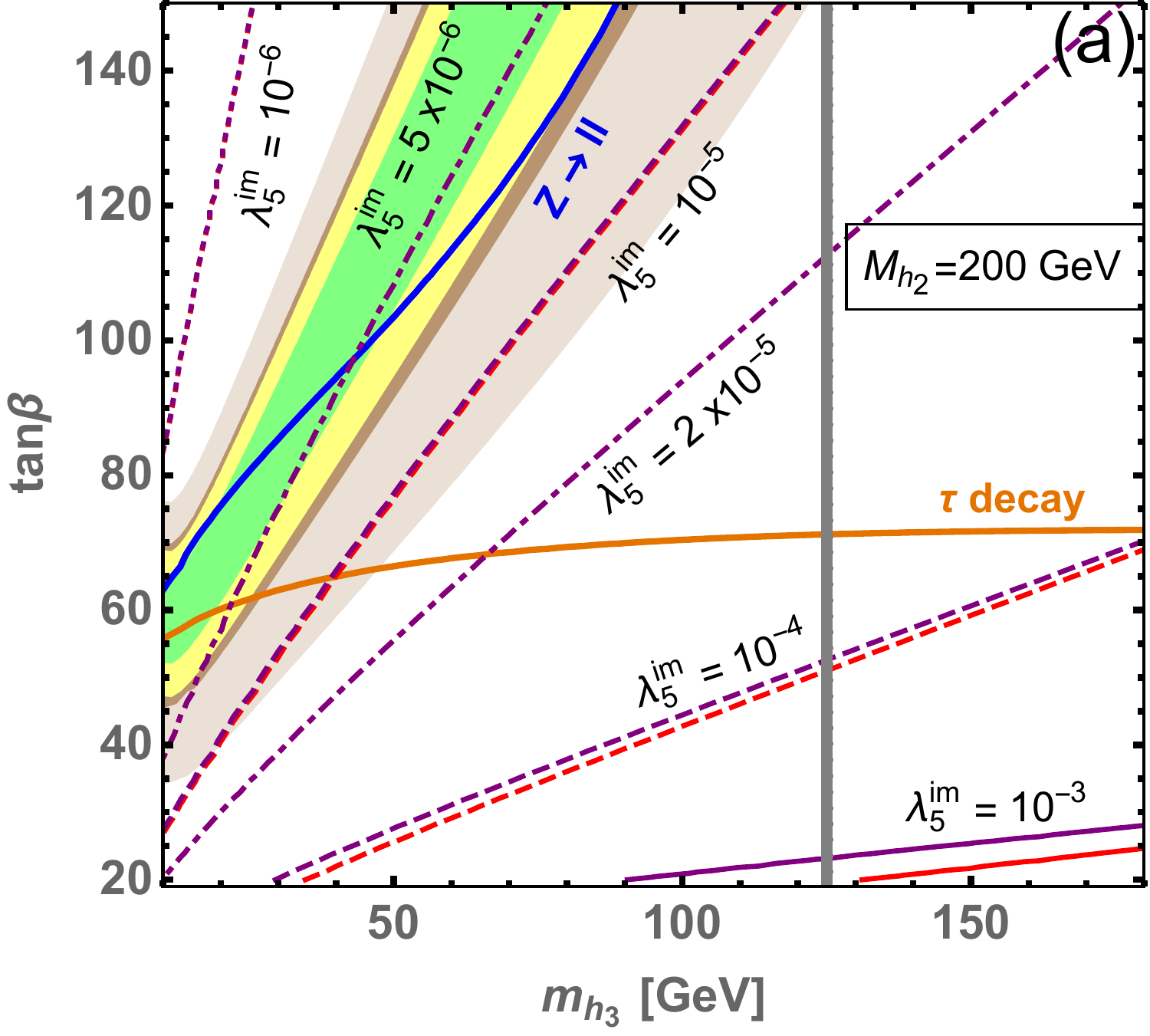}
  \includegraphics[width=7.2cm,angle=0]{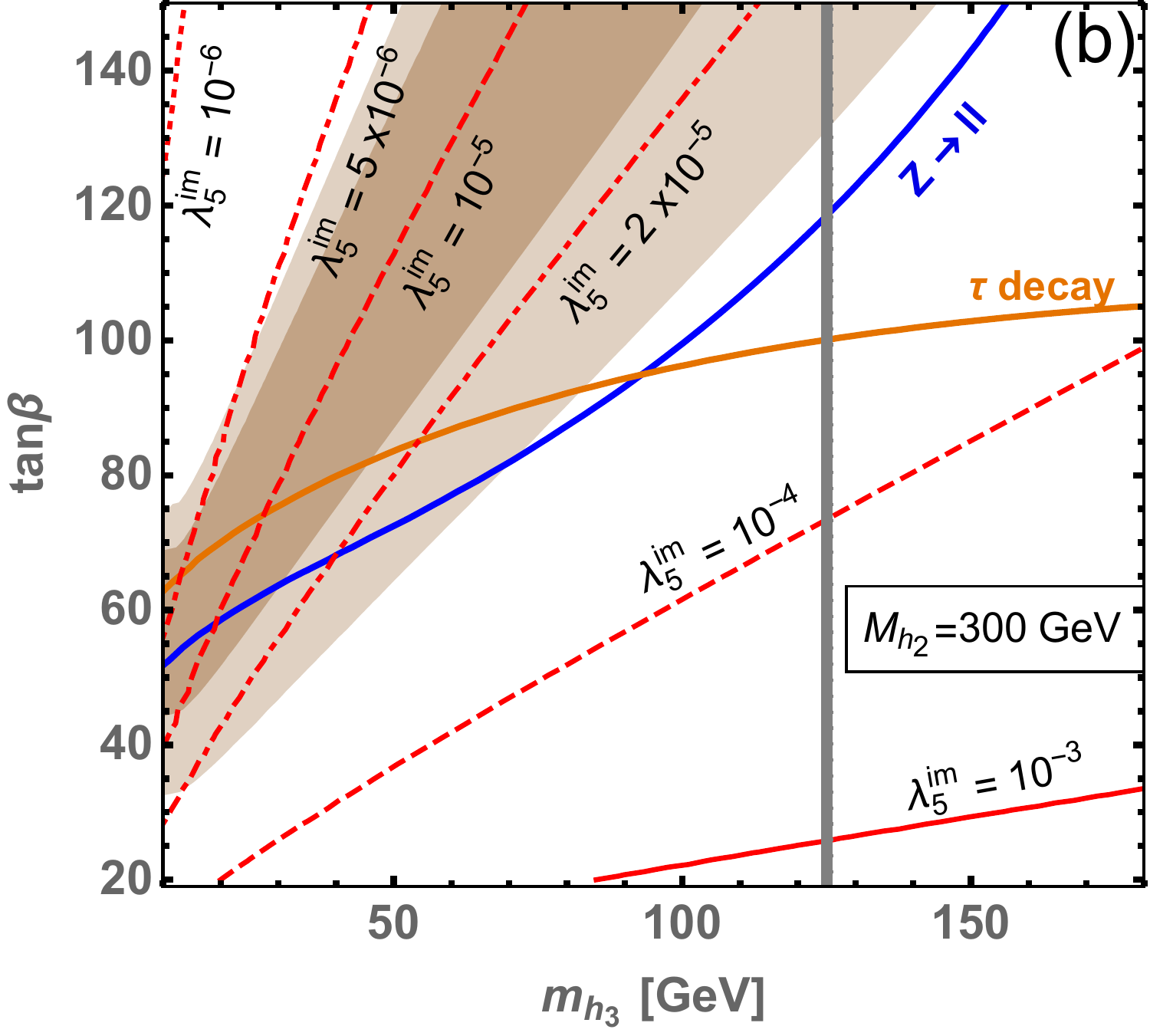}
  \includegraphics[width=7.2cm,angle=0]{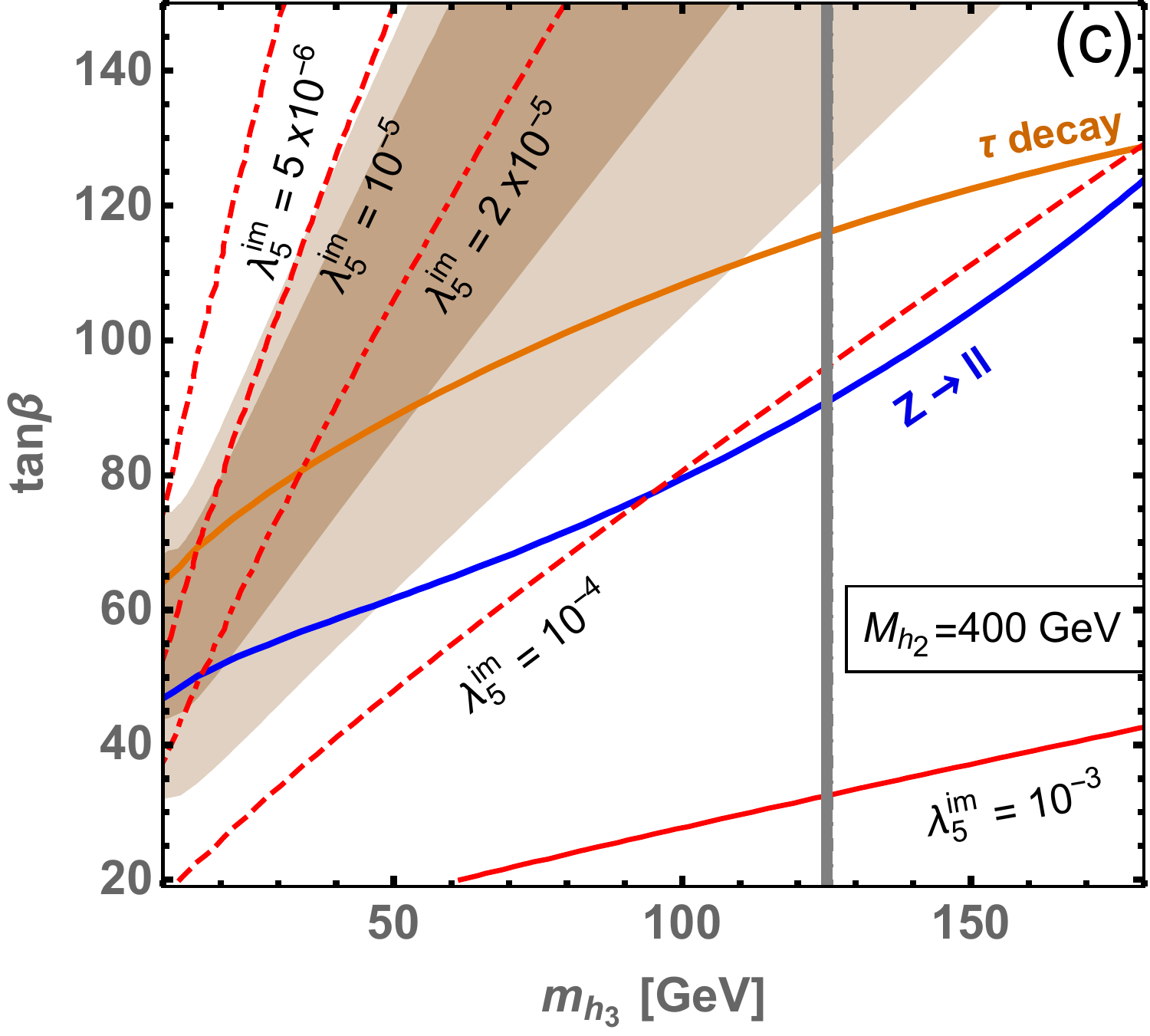}  
  \caption{Allowed parameter space in $m_{h_3}-\tan\beta$ plane depicting all the constraints coming from EDM, $(g-2)_\mu$ and lepton flavor universality. The parameter ${\rm Im}(\lambda_5)$ is denoted as  $\la_5^{\rm{im}}$ in the plots. }
 \label{fig:ma_tanb}
\end{center}
\end{figure}
In $m_{h_3}-\tan\beta$ plane we present our main result where we show the parameter space compatible with $(g-2)_\mu$, electron EDM and lepton universality 
constraints coming from $Z\to\ell\ell$ and $\tau$ decay measurements. In Fig.~\ref{fig:ma_tanb}(a) the light and dark brown region can explain the present $(g-2)_\mu$ anomaly (Eq.~\ref{eq:delta-amu}) at $1\sigma$ and $2\sigma$ respectively. The region in green(yellow) depicts parameter space which will be able the explain the anomaly at $1\sigma(2\sigma)$ after full Fermilab data assuming that the central values remain the same. The parameter space right or below to the dark blue and orange lines are  allowed at $2\sigma$ 
by lepton universality in $Z$ decays and $\tau$ decays. The constraints coming from the electron EDM is shown by the red and purple curves which depends on the value of CPV coupling  ${\rm Im}(\lambda_5)$. For a given ${\rm Im}(\lambda_5)$, along the red or purple curve the electron EDM is $1.1\times 10^{-29} $ e-cm and anything right to that curve is allowed by the present EDM limit. For a light $m_{h_3}$ the BZ contribution is very large and consequently a very small CPV coupling is allowed. On the other hand when $h_3$ is relatively heavy, say 150 GeV then relatively large  ${\rm Im}(\lambda_5)$ is allowed. Similarly for large $\tan\beta$ the EDM contribution enhanced by $\tan^2\beta$ and for a given $m_{h_3}$ if we increase $\tan\beta$ the limit on CPV coupling becomes stronger. The red(purple) curves in Fig.~\ref{fig:ma_tanb}(a) are the EDM constraints for WS (RS) Yukawa limit since for $m_{h_2}=200$ GeV both the RS and WS Yukawa couplings are allowed. The vertical gray bar in Fig.~\ref{fig:ma_tanb} depicts the region where $h_3$ and the the SM Higgs $h_1$ becomes degenerate and the 
approximate expressions for mixing angles as in Eq.~\ref{eq:mixing_angles} breaks down. 
In  Fig.~\ref{fig:ma_tanb}(b),(c) we have shown the allowed parameter space for $m_{h_2} = m_h^\pm =300$ and 400 GeV and for simplicity the future Fermilab 
limit is not shown. The light(dark) brown region explain the present $(g-2)_\mu$ anomaly at 2~$\sigma$(1~$\sigma$). As we increase $m_{h_2}$ the mixing angle $\sin\alpha_c$ decreases(see Eq.~\ref{eq:mixing_angles}) and comparatively large ${\rm Im}(\lambda_5)$ is allowed for a given 
$m_{h_3}$ and $\tan\beta$. 
Note that for $m_{h_2} \geq 250$ GeV, RS Yukawa couplings are disfavored because they cannot satisfy the 
perturbativity and vacuum stability bounds as in Eq.~\ref{eq:vacpert}. Hence for heavier $m_{h_2}$ we have only WS Yukawa limit.
\section{Conclusion and remarks} \label{conclusions}

We explored the parameter space of the CP violating type-X 2HDM in which sizable enhancement of electric and magnetic moment of leptons can arise through two-loop Barr-Zee diagrams. Figure \ref{fig:ma_tanb} summarizes our main results on the limits from the muon 
$g-2$, electron EDM, and lepton universality determinations in the plane of $( m_{h_3},\tan\beta)$. 
For this, we imposed the theoretical constraints of vacuum stability and perturbativity, and assumed degenerate masses for the charged Higgs boson and the heaviest neutral Higgs boson to be consistent with the electroweak precision test.

Let us remark that the region explaining the muon $g-2$ anomaly is more tightly constrained by the lepton universality conditions 
compared with the previous studies, e.g., in \cite{jinsu16}. 
This is because we used the new theoretical value \cite{Keshavarzi:2018mgv} which increased the deviation a bit
 (Eq.~\ref{eq:delta-amu}). Future collider experiments improving the precision of  lepton universality would be 
useful to test the type-X 2HDM as an explanation to the muon $g-2$ deviation.
In the parameter region explaining the muon $g-2$ anomaly, the electron EDM is also enhanced and thus one can see that 
the CPV quartic coupling has to be smaller than about  a few times $ 10^{-5}$ to be compatible with the muon $g-2$ explanation.

\appendix

\section{Wilson coefficients for EDM in 2HDM}\label{app:wilson}
All the necessary Wilson coefficients for EDM are given in this appendix. 
\subsection{Diagrams with fermion loop}
\begin{itemize}
 \item $h_i\gamma\gamma$ diagram : 
\be\label{A6}
\left(\delta_e \right)^{h_i\gamma\gamma}_{f} =  N_c\  Q_{f}^2\ e^2 \frac{1}{64\pi^4} \sum_{i=1}^3 \left[ f(z^i_{f})\ c_{f,i} \tilde c_{e,i} + g(z^i_{f})\ \tilde c_{f,i} c_{e,i} \right] \ ,
\ee
where $f = \tau,b,t$. The loop functions $f(z^i_{f})$ and $ g(z^i_{f})$ are written in Appendix~\ref{app:loop_fn}.
\item $h_i Z\gamma$ diagram:
\be\label{A7}
\left(\delta_e \right)^{h_iZ\gamma}_{f} = - N_c  g_{Z\bar e e}\ g_{Z\bar f f}\ \frac{1}{64\pi^4} \sum_{i=1}^3 \left[ \tilde f(z^i_{f}, m_{f}^2/M_Z^2) c_{f,i} \tilde c_{e,i} + \tilde g(z^i_{f}, m_{f}^2/M_Z^2) \tilde c_{f,i} c_{e,i} \right] \ ,
\ee
with 
$g_{Zf\bar f} = g(T_3^f - 2 Q^f \sin^2\theta_W)/({2\cos\theta_W}) $. The loop functions $\tilde f$ and $\tilde g$ are in Appendix~\ref{app:loop_fn}. 
\end{itemize}
\subsection{Diagrams with Charged Higgs or W boson loop}
\begin{itemize}
 \item Diagrams with $W^\pm$ boson :
 \begin{eqnarray}\label{gammaW}
\left(\delta_e \right)^{h\gamma\gamma}_W &=& - e^2 \frac{1}{256\pi^4} \sum_{i=1}^3 \left[ \left( 6 + \frac{1}{z^i_w} \right) f(z^i_w) + \left( 10 - \frac{1}{z^i_w} \right) g(z^i_w) \right] a_i \tilde c_{e,i}  \ , \label{A8} \\
\left(\delta_e \right)^{hZ\gamma}_W &=& g_{Z\bar e e} g_{ZWW} \frac{1}{256\pi^4} \sum_{i=1}^3 \left[ \left(6 -\sec^2\theta_W + \frac{2-\sec^2\theta_W}{2z^i_w} \right)\tilde f(z^i_w, \cos^2\theta_W) \right.\nonumber \\
&&\hspace{1cm}+ \left. \left( 10- 3\sec^2\theta_W - \frac{2-\sec^2\theta_W}{2z^i_w}\right)\tilde g(z^i_w, \cos^2\theta_W) \right] a_i \tilde c_{e_i} \ , \label{A9}
\end{eqnarray}
where $z^i_w = M_m^2/m_{h_i}^2$ and $g_{WWZ}/e = \cot\theta_W$.
\item Diagrams with $h^\pm$ :
\begin{eqnarray}\label{gammaH+}
\left(\delta_e \right)^{h\gamma\gamma}_{h^+} &=& - e^2 \frac{1}{256\pi^4} \left(\frac{v}{m_{h^+}}\right)^2 \sum_{i=1}^3\left[ f(z^i_h) - g(z^i_h)\right] \bar \lambda_i \tilde c_{e,i} \ ,\label{A10} \\
\left(\delta_f \right)^{hZ\gamma}_{h^+} &=& g_{Z\bar e e} g_{Zh^+ h^-} \frac{1}{256\pi^4} \left( \frac{v}{m_{h^+}}\right)^2 \sum_{i=1}^3 \left[ \tilde f(z^i_H, m_{h^+}^2/M_Z^2) - \tilde g(z^i_h, m_{h^+}^2/M_Z^2)\right] \bar \lambda_i \tilde c_{e,i} \ ,\nonumber  \\\label{A11}
\end{eqnarray}
with $z^i_H = m_{h^+}^2/m_{h_1}^2$ and $g_{Zh^+h^-}/e = \cot\theta_W(1-\tan^2\theta_W)/2$.
\end{itemize}
\subsection{Diagrams with $h^\pm W^\mp \gamma$}
These contributions are taken from ref.~\cite{Abe:2013qla}.
\be
\left(\delta_e \right)^{h^\pm W^\mp\gamma} = \frac{\mathcal{S}_l}{512\pi^4}  \sum_i \left[ \frac{e^2}{2\sin^2\theta_W} \mathcal{I}_4(m_{h_i}^2, m_{h^+}^2) a_i \tilde c_{e,i} - \mathcal{I}_5(m_{h_i}^2, m_{h^+}^2) g_{h_i h^\pm h^\mp}\ \tilde c_{e,i} \right],
\ee
where $\mathcal{I}_{4,5}$ is given in Appendix~\ref{app:loop_fn}. The  $\mathcal{S}_l = +1$ for charged lepton and down type quarks and  
-1 for up type quarks.
\section{Loop Functions}\label{app:loop_fn}
\bea
f(z) &=& \frac{z}{2} \int_0^1 dx \frac{1-2x(1-x)}{x(1-x)-z} \textrm{ln}\left(\frac{x(1-x)}{z}\right) \ , \\
g(z) &=& \frac{z}{2} \int_0^1 dx \frac{1}{x(1-x)-z} \textrm{ln}\left(\frac{x(1-x)}{z}\right) \ , \\
\tilde f(x,y) &=& \frac{y\ f(x)}{y-x} + \frac{x\ f(y)}{x-y}  \ , \\
\tilde g(x,y) &=& \frac{y\ g(x)}{y-x} + \frac{x\ g(y)}{x-y} \ , \\
\mathcal{I}_{4,5}(m_1^2, m_2^2) &=& \frac{m_W^2}{m_{h^+}^2 - m_W^2} \left(I_{4,5}(m_W^2, m_1^2)-I_{4,5}(m_2^2, m_1^2)\right) \ , \\
I_4(m_1^2,m_2^2) &=& \int_0^1 dz (1-z)^2 \left( z-4 + z\frac{m_{h^+}^2 -m_2^2}{m_W^2} \right)  \nonumber \\
&&\hspace{0.2cm} \times \frac{m_1^2}{m_W^2(1-z)+m_2^2 z - m_1^2z(1-z)} \textrm{ln}\left(\frac{m_W^2(1-z)+m_2^2z}{m_1^2z(1-z)}\right) , \   \ \\
I_5(m_1^2,m_2^2) &=& \int_0^1 dz\ \frac{2\ m_1^2 z(1-z)^2}{m_{h^\pm}^2(1-z)+m_2^2 z - m_1^2z(1-z)} \textrm{ln}\left(\frac{m_{h^\pm}^2(1-z)+m_2^2z}{m_1^2z(1-z)}\right)\   \ \nonumber\\
\eea


\end{document}